\documentclass[reprint,
preprintnumbers,amsmath,amssymb]{revtex4}

\usepackage{amsmath,amsthm}
\usepackage{amssymb,latexsym}
\usepackage{graphicx}
\usepackage{setspace}
\usepackage[mathscr]{eucal}
\usepackage{subfig}
\usepackage{hyperref}

\numberwithin{equation}{section}

\begin{document}
\title{Spatiotemporal two-dimensional solitons in the complex Ginzburg-Landau equation}

 \author{Florent B\'erard}
 \email{fberard@enseirb-matmeca.fr}
 \affiliation{Department of Mathematics, ENSEIRB-MATMECA, Universit\'e Bordeaux 1, France 33405}

\author{Stefan C. Mancas}
\email{mancass@erau.edu}
\affiliation{Department of Mathematics, Embry-Riddle Aeronautical University, Daytona Beach, FL 32114-3900, USA}

\begin{abstract}
We introduce spatiotemporal solitons of the two-dimensional complex Ginzburg-Landau equation (2D CCQGLE) with cubic and quintic nonlinearities in which asymmetry between space-time variables is included. The 2D CCQGLE is solved by a powerful Fourier spectral method, i.e., a Fourier spatial discretization and an explicit scheme for time differencing. Varying the system's parameters, and using different initial conditions, numerical simulations reveal 2D solitons in the form of stationary, pulsating and exploding solitons which possess very distinctive properties. For certain regions of parameters, we have also found stable coherent structures in the form of spinning (vortex) solitons which exist as a result of a competition between focusing nonlinearities and spreading while propagating through medium. 
\end{abstract}

\maketitle
\section{Introduction}
The complex cubic-quintic Ginzburg-Landau equation (CCQGLE) is the canonical equation governing the weakly nonlinear behavior of dissipative systems in a wide variety of disciplines \cite{Dodd}. In fluid mechanics, it is also often referred to as the Newell--Whitehead equation after the authors who derived it in the context of B\'enard convection \cite{Dodd,Drazin:1}. As such, it is also one of the most widely studied nonlinear equations. Many basic properties of the equation and its solutions are reviewed in \cite{Aranson,Bowman}, together with applications to a vast variety of phenomena including nonlinear waves, superconductivity, Bose--Einstein condensation, liquid crystals and string theory. In particular, in nonlinear optics it describes the pulse generation and signal transmission through an optical fiber.

An important element  in the long time dynamics of pattern forming systems is a class of solutions which we call ``coherent structures" or solitons. These structures could be a profile of light intensity, temperature, magnetic field, etc. A dissipative soliton is localized and exists for an extended period of time. Its parts are experiencing gain/loss of energy or mass with the medium. 

In contrast to the regular solitary waves investigated in numerous integrable and non-integrable systems over the last three decades, these dissipative solitons are not stationary in time \cite{Akhmediev:7}. Rather, they are spatially confined pulse-type structures whose envelopes exhibit complicated temporal dynamics. Another relevant feature of dissipative systems is that they include energy exchange with external sources. Thus, energy can flow into the system through its boundaries. As long as the parameters in the system stay constant, the structure evolves by changing shape and exists indefinitely in time. The structure disappears when the source is switched off, or if the parameters are moved outside of the range of existence of the soliton solutions. These soliton solutions appear as a result of balance between diffraction (dispersion) and nonlinearity. Diffraction spreads a beam while nonlinearity will focus it and make it narrower. The balance between the two results in stationary solitary wave solutions, which usually form a one parameter family. In dissipative systems with gain and loss, in order to have stationary solutions, the gain and loss must be also balanced. This additional balance results in solutions which are fixed. Then the shape, amplitude and the width are all completely fixed by the parameters of the dissipative equation. However, the solitons, when they exist, can again be considered as ``modes" of dissipative systems just as for non-dissipative ones \cite{Akhmediev:7, Akhmediev:3, Akhbook}. 

When a system is integrable, its solutions can be constructed using the inverse scattering technique \cite{Ablowitz}. Further generalization
to non-integrable systems shows that traveling wave solutions can be found and analyzed using some
basic approaches from nonlinear dynamics. One of the powerful
techniques that can be applied to these systems is that which
employs Hamiltonian equations. For dissipative systems, there are no conserved quantities that
can be used to solve particular problems.

Recent perturbative treatments based on expressions about the nonlinear Schr\"odinger equations are generalized to perturbations
of the cubic-quintic and derivative Schr\"odinger equations. The cubic Ginzburg-Landau admits a selected range of exact soliton
solutions \cite{Maruno}. Even the known special solutions have exotic dynamic responses to external conditions, so that sensitivity analysis of the parameters is paramount. These exist when certain relations between parameters are satisfied. However, this certainly does not imply that the equations are integrable. In reality, general dissipative nonlinear PDEs cannot be reduced to linear equations in any known way, therefore an insight to the type of solutions that the equation has may be based on numerical simulations. To investigate numerically the two dimensional dissipative soliton solutions of the CCQGLE, and analyze their qualitative behavior, we used a powerful Fourier spectral method. 
 
\section {2D CCQGLE}

The 2D CCQGLE is solved by spectral methods, i.e., a Fourier spatial discretization and an explicit scheme for time differencing.  In addition, it is also crucial to analyze the dependence of both their shape and stability on the various parameters of the CCQGLE,
viz. the nonlinearity, dispersion, linear and nonlinear gain, loss and spectral filtering parameters of these new classes of dissipative solitons, since the solutions experience interesting bifurcation sequences as the parameters of the CCQGLE are varied. A rich variety of  evolution behaviors, which include pulsating, exploding and spinning dissipative soliton dynamics is revealed. 

In dimensionless form, the 2D CCQGLE with the corresponding cubic-quintic terms is a modified NLS and it takes the form \cite{Ankiewicz:2,Akhmediev:7}
\begin{equation}\label{2.1}
\partial_tA=\epsilon A+(b_1+ic_1)\nabla_{\bot}^2 A-(b_3-ic_3) |A|^2A-(b_5-ic_5) |A|^4 A,
\end{equation}
where $\nabla_{\bot}^2 =\frac 1 r \frac{\partial}{\partial r}\big(r\frac{\partial}{\partial r}\big)$ is the transverse Laplacian for radially symmetric beams. The present study will confine itself to the spatially infinite system in two dimensions  and will focus primarily on a spatiotemporal behavior of dissipative solitons without imposing radial symmetry on the beams, and hence the Laplacian $\nabla_{\bot}^2 =\frac{\partial^2}{\partial x^2}+\frac{\partial^2}{\partial y^2}$ depends on both variables $x$ and $y$.

The interpretation of the system's parameters in (\ref{2.1}) depends on the particular field of work. In optics, $A(r;t)$ is the normalized envelope of the field, $r$ is the radial coordinate, $t$ is the propagation distance or the cavity number.  The system's parameters of (\ref{2.1}) are: $\epsilon$ linear loss/gain, $b_1$- angular spectral filtering, $c_1=0.5$- second-order diffraction coefficient, $b_3$- nonlinear gain/loss, $c_3=1$- nonlinear dispersion, $b_5$- saturation of the nonlinear gain/loss, and $c_5$- saturation of the nonlinear refractive index. 

\section{Fourier Spectral Method}

The Fourier transform of a function $u(x,y)$ is defined by
\begin{equation}\label{four}
\mathcal{F}(u)(k_x,k_y)=\widehat{u}(k_x,k_y)=\frac{1}{2\pi}\int_{-\infty}^{\infty}\int_{-\infty}^{\infty}e^{-i(k_xx+k_yy)}u(x,y)\, dxdy,
\end {equation}
with the corresponding inverse
\begin{equation}
\mathcal{F}^{-1}(\widehat{u})(x,y)=u(x,y)=\frac{1}{2\pi}\int_{-\infty}^{\infty}\int_{-\infty}^{\infty}e^{i(k_xx+k_yy)}\widehat{u}(k_x,k_y)\, dk_xdk_y.
\end {equation}

The function $\widehat{u}(k_x,k_y)$ can be interpreted as the amplitude density of $u$ for wavenumbers $k_x$, $k_y$. 

First, we write the Fourier transform of equation (\ref{2.1}) as
\begin{equation}\label{a}
\widehat{A_t}=\left[\epsilon-(b_1+ic_1)(k_x^2+k_y^2)\right]\widehat{A}-(b_3-ic_3)\widehat{|A|^2A }-(b_5-ic_5)\widehat{|A|^4A},
\end{equation}
and rearranging the terms in (\ref{a}), we consider the following ordinary differential equation in the Fourier space

\begin{equation}\label{b}
\widehat{A_t}=\alpha (k_x,k_y)\widehat{A}+\beta \widehat{|A|^2A }+\gamma \widehat{|A|^4A},
\end{equation}
with $\alpha (k_x,k_y)= \epsilon-(b_1+ic_1)(k_x^2+k_y^2)$, $\beta=-(b_3-ic_3)$ and $\gamma=-(b_5-ic_5)$.
In the Fourier space, (\ref{b}) contains the linear part $\alpha (k_x,k_y)\widehat{A}$, and the non-linear part $\beta \widehat{|A|^2A }+\gamma \widehat{|A|^4A}$. 

We solve the initial value problem for the above ODE  numerically using an explicit scheme  i.e., $4^\textrm{th}$ order Adams-Bashforth for the non-linear part and the exact solution for the linear part, which is given by
\begin{equation}\label{Exact}
\widehat{A}(t)=\widehat{A(x,y;0)}e^{\alpha (k_x,k_y)t}.
\end{equation}

\subsection{Spatial discretization (Discrete Fourier Transform)}
\noindent
We use a standard discretization of a rectangular spatial domain $[-L/2,L/2] \times [-L/2,L/2]$ into $n \times n$ uniformly spaced grid points $X_{ij}=(x_i,y_j)$ with $\Delta x=\Delta y= L/n$ and $n$ even. Given $A(X_{ij})=A_{ij},\, i,j=1,2,\cdots,n$, we define the 2D Discrete Fourier transform (2DFT) of $A$ as
\begin{equation}\label{eq41}
\widehat{A}_{k_xk_y} = \Delta x \Delta y \sum_{i=1}^{n} \sum_{j=1}^{n} e^{-i(k_xx_i+k_yy_j)}A_{ij},\,\,\,\,\,\, k_x,k_y=-\frac{}{2}+1,\cdots, \frac{n}{2}
\end{equation}
and its inverse 2DFT as
\begin{equation}\label{eq5}
A_{ij} = \frac{1}{(2\pi)^2} \sum_{k_x=-n/2+1}^{n/2} \sum_{k_y=-n/2+1}^{n/2} e^{i(k_xx_i+k_yy_j)}\widehat{A}_{k_xk_y},\,\,\,\,\,\, i,j=1,2,\cdots, n.
\end{equation}
In (\ref{eq41})-(\ref{eq5}) the wavenumbers $k_x$ and $k_y$, and the spatial indexes $i$ and $j$,
take only integer values.

\subsection {Temporal discretization ($4^\textrm{th}$ order Adams-Bashforth)}

Given $t_\textrm{max}$ we discretize the time domain $[0,t_\textrm{max}]$ with equal time steps of width $\Delta t$ as  $t_n=n\Delta t , \, n=0,1,2,\cdots,$ and define
$\widehat{A}^n=\widehat{A}(t_n)$.
Initializing $\widehat{A}^n=\widehat{A}(t_n)$, we compute the nonlinear terms
$\mathcal{N}_3=\mathcal{F}\left(\left|\mathcal{F}^{-1}(\widehat{A}^n)\right|^2\mathcal{F}^{-1}(\widehat{A}^n)\right)$, and $\mathcal{N}_5=\mathcal{F}\left(\left|\mathcal{F}^{-1}(\widehat{A}^n)\right|^4\mathcal{F}^{-1}(\widehat{A}^n)\right)$
and advance the ODE (\ref{b}) in time
with time step $\Delta t$ using the 4-step Adams-Bashforth method. 

The discretized ODE has the following form
\begin{equation}
\widehat{A}^{n+1}=\widehat{A}^{n}e^{\alpha(k_x,k_y)t}+\frac{\Delta t}{24}\left[55f(\widehat{A}^{n})-59f(\widehat{A}^{n-1})+37f(\widehat{A}^{n-2})-9f(\widehat{A}^{n-3})\right],
\end{equation}
with exact solution of the linear part $\widehat{A}^{n+1}=\widehat{A}^{n}e^{\alpha(k_x,k_y)t}$,
and the nonlinear part $f$ defined by $f(\widehat{A})=\beta\mathcal{N}_1 +\gamma \mathcal{N}_2$.

\subsection{Numerical implementation}\label{Numerical method}
\noindent
The language used is Fortran 90, with gfortran compiler. All the parameters needed by the program are written in a datafile so that they can be easily modified, without systematic compilation. Among them, most important are the parameters of the  CCQGLE, which are shown in Table \ref{TabParam}.  
For the ring solitons from \textsl{\S 4.4} we have used the 1D case of the parameters of \cite{Akhmediev:1}, while for the stationary and pulsating, the parameters are from \cite{Soto:2,Ankiewicz:2}. For the exploding soliton we have used the parameters from  \cite{Soto:3,Crespo:1}. All the parameters of the 1D case are extended in this paper to the 2D case.

\begin{table}
\begin{center}
\begin{tabular}{|c|ccccccc|}
\hline 2D solitons & $\epsilon$ & $b_1$ & $c_1$ & $b_3$ & $c_3$ & $b_5$ & $c_5$\\
\hline
stationary & -0.045 & 0.04 & 0.5 & -0.21 & 1 & 0.03 & -0.08  \\
\hline
ring & -0.1 & 0.1 & 0.5 & -0.8 & 1 & 0.04 & -0.02  \\
\hline
pulsating & -0.045 & 0.04 & 0.5 & 0.37 & 1 & 0.05 & -0.08 \\
\hline
exploding & -0.1 & 0.125 & 0.5 & -1 & 1 & 0.1 & -0.6 \\
\hline
\end{tabular}
\end{center}
\caption{Sets of parameters for 2D solitons}
\label{TabParam}
\end{table}

An important subroutine is the Fast Fourier Transform (FFT) \cite{f90Recipes,Matlab}. This subroutine uses an algorithm that reduces computational time from $\mathcal{O}(n^2)$ to  $\mathcal{O}(n\log n)$ floating point operations. Moreover, it only works for periodic functions hence, caution must be made when we select the size of the domain for different cases of the soliton solutions. This must be chosen in such way that the solitons do not touch the boundaries, otherwise if the non-zero part of the solutions propagates through some boundary, it will spread throughout  the domain from the opposite side. 

The Fortran program ran on the Zeus cluster at Embry-Riddle Aeronautical University. This parallel computer of 256 nodes is rated to over 1 Tera-flop. Jobs were submitted thanks to Platform Lava installed on the cluster. This software manages to distribute all the jobs to different nodes, considering priorities. Once the output file are written, we used Tecplot 360 for visualization.

The duration of computations depends on the equation parameters and mesh size. The most appropriate size to use was 512x512, but other types of meshes were considered. Very fast calculations took almost two hours, while longer ones took several days.

\section{Numerical Simulations}
\subsection{Initial conditions}
\begin{figure}[htbp]
\centering
\includegraphics[width=350pt]{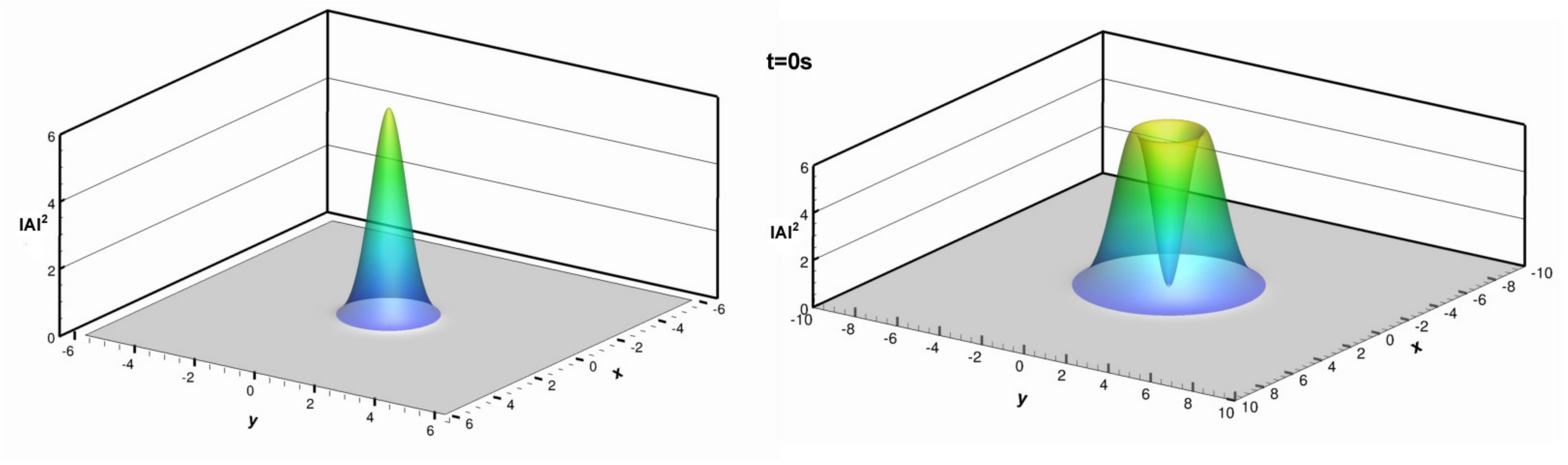}
\caption{Initial shapes of the field. Left: Gaussian shape. Right: Ring shape with vorticity $m=1$.}
\label{Initial}
\end{figure}

Localized structures with vorticity are obtained when initial conditions have radial symmetry, and hence we assume that the initial pulse can be reasonably well approximated by a bell-shaped, i.e.,
\begin{itemize}
\item[i.] Gaussian shape,
\begin{equation}
A(x,y;0)=A_0e^{-r^2}
\end{equation}
\item[ii.] Ring shape with rotating phase,
\begin{equation}
A(x,y;0)=A_0r^me^{-r^2}e^{im\theta}
\end{equation}
where $m$ is the degree of vorticity, $A_0$ is a real amplitude that should generate sufficient power to place the initial condition into a basin of attraction of a 2D soliton, and $\theta=\tan^{-1}{\big(\frac{\sigma_yy}{\sigma_xx}\big)}$ is the phase. The widths of each of the initial conditions could be either circular (radially symmetric) or elliptic, and are controlled by the inverse widths parameters $\sigma_x$ and $\sigma_y$, with $r=\sqrt{(\sigma_xx)^2+(\sigma_yy)^2}$, see Fig. \ref{Initial}. In this paper we have used the vorticity $m=1$, but family of solitons of higher vorticity have been found \cite{Soto:4}.
\end{itemize}

\subsection{Parameters}

The existence of stable solutions is strongly linked to the parameters of the equation. One of the most important issue is to find the continuous range of existence of solitons of the same type in parameters space. Moreover, several solutions can coexist for the same set of parameters. These solitons are not necessarily stable. Hence, stability is controlled by the parameters of the equation and by the choice of the initial conditions.

The solutions will be computed using a  Fourier spectral method in which we will monitor the energy $Q(z)=\int_{-\infty}^{\infty}\int_{-\infty}^{\infty} |A(x,y;z)|^2 \, dxdy$ \cite{Soto:2}.
For a localized solution, $Q$ is finite and changes smoothly while the solution stays within the region of existence of the soliton \cite{Crespo:1}. When $Q$ changes abruptly there is a bifurcation and the solution jumps from a branch of solitons that become unstable to another branch of stable solitons, or vice versa. As soon as the solution becomes unstable, $Q$ diverges until infinity or collapses to $0$. For a certain class of solutions, $Q$ will evolve periodically in some regime, and will converge to a finite value \cite{Crespo:1, Soto:2}.

\subsection{Stationary solitons}\label{Stationary}

The initial shape is a circular 2D Gaussian with $A_0=2.5$, and $\sigma_x=\sigma_y=1$.
Steady solitons are the simplest solutions we can find. After a transition period of stabilization, the shape of the solitary waves remains the same. The energy becomes 
constant after $z\approx40 s$. The soliton profile is always a Gaussian, and each profile
has a plain bell-shape.
\begin{figure}[htbp]
\centering
\includegraphics[width=250pt]{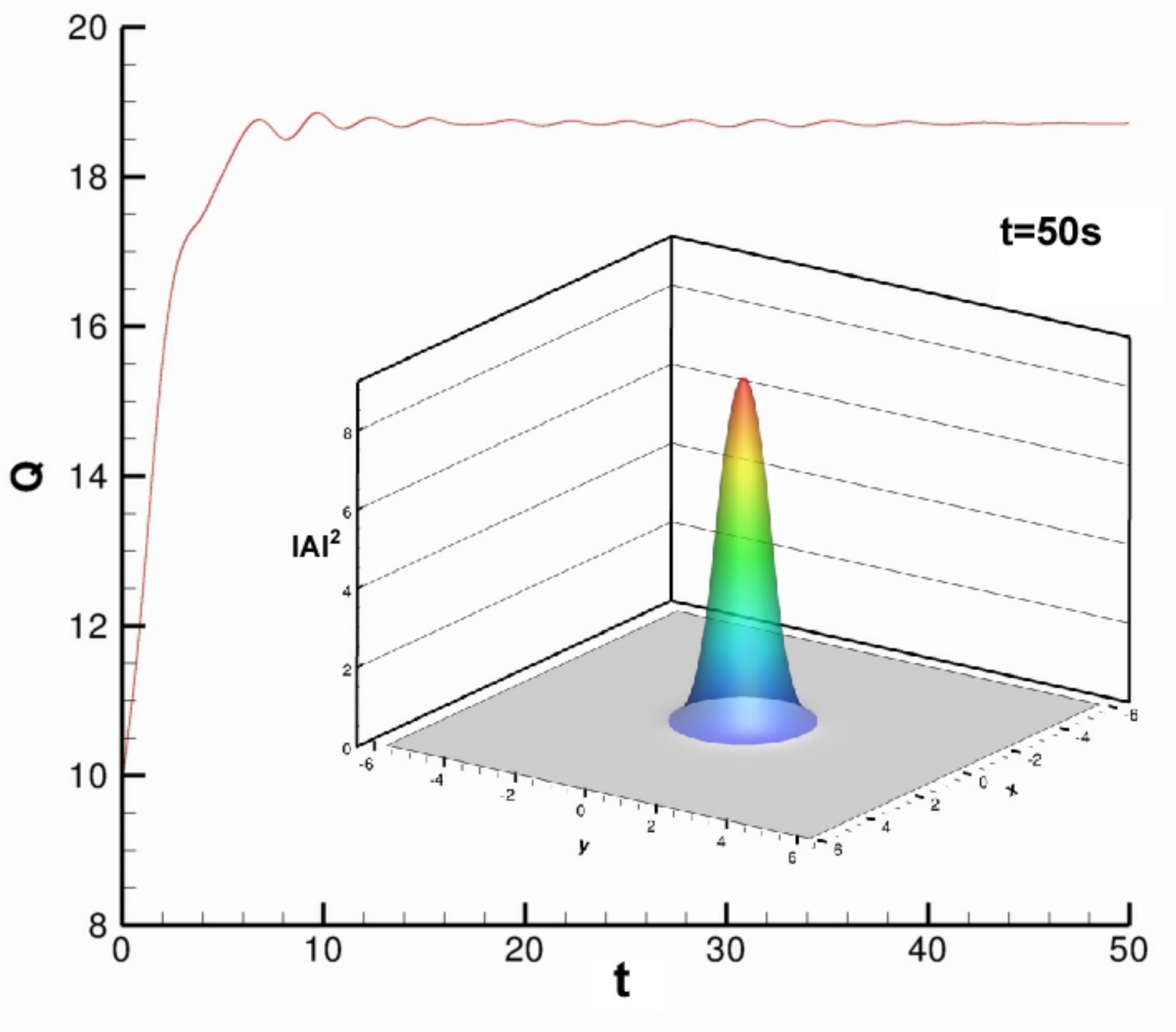}
\caption{Stationary soliton. Energy is concentrated in the center of the domain, in a bell-shaped structure. Parameters: line 1 of Table \ref{TabParam}.}
\label{Normal}
\end{figure}
Finally, energy converges to $Q(t) \approx18.6$, and stabilizes forever if the system's parameters are not changed.
\subsection{Ring vortex (spinning) solitons}
\begin{figure}[htbp]
\centering
\includegraphics[width=350pt]{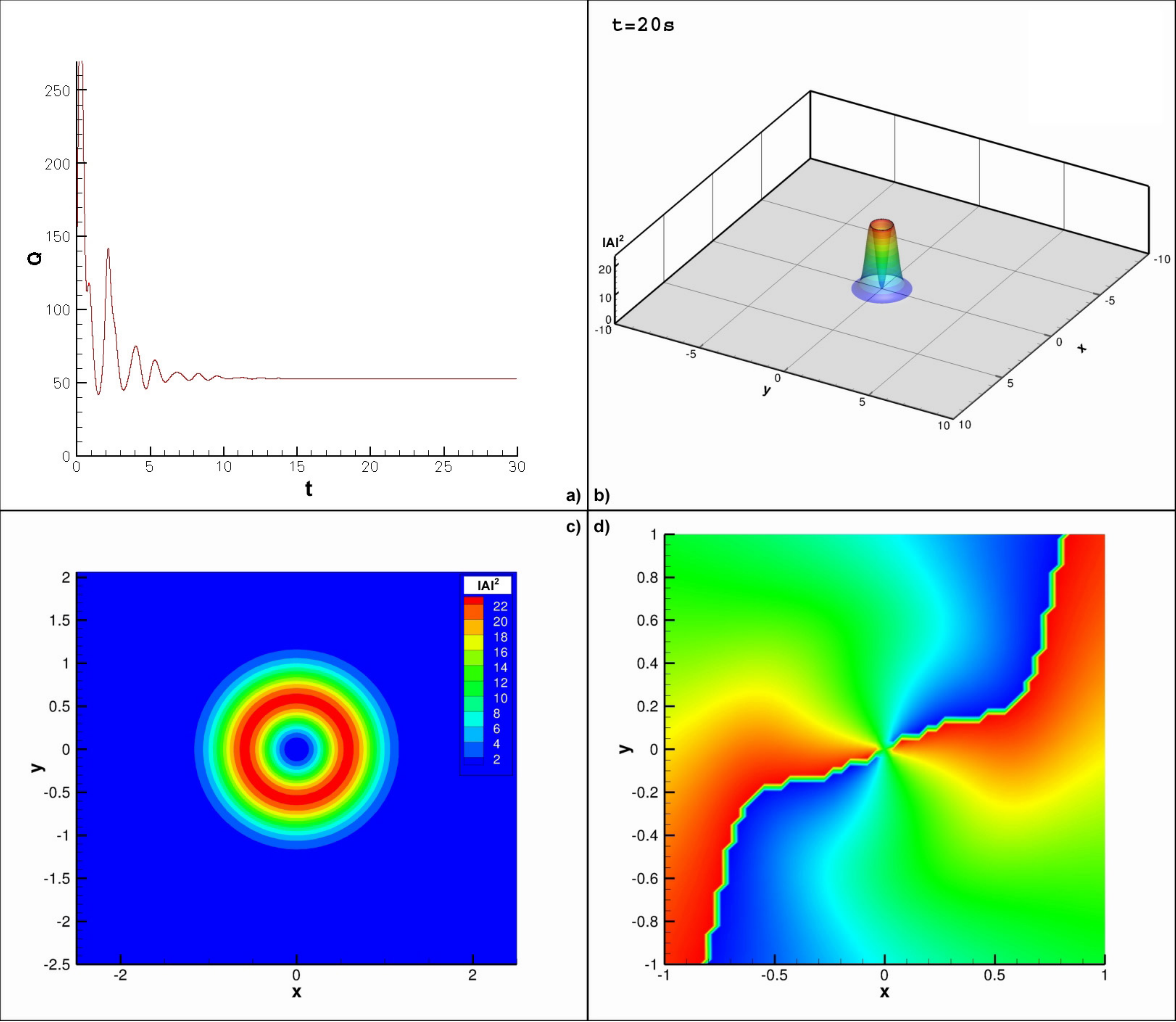}
\caption{Circular ring vortex soliton with radially symmetry. Top left: Energy. Top right: Ring vortex soliton at $t=20s$. Bottom left: Contour plot of $|A|^2$. Bottom right: Phase plot that indicates rotation at $t=20 s$.  Parameters: line 2 of Table \ref{TabParam}.}
\label{Ring}
\end{figure}

We choose a circular ring type structure with vorticity $m=1$, $A_0=2.5$, and $\sigma_x=\sigma_y=1$, and parameters from line 2 of Table \ref{TabParam}. By keeping all the parameters fixed, we can find different types of ring vortex solitons, by varying the initial conditions. This also can be achieved by varying slowly one of the equation parameters while the initial condition stays fixed. First, we found a radially symmetric ring vortex soliton shown in Fig. \ref{Ring}. Even though the structure oscillates before converging to a stable value of the energy as it is with the stable soliton, now since we introduced a phase, the soliton is spinning. The rotation has a fixed angular velocity and can occur in either direction clockwise or counterclockwise, which is determined by the initial condition.  

Next, we introduced for the same set of parameters, a different initial condition, viz. an elliptic shape which breaks the  symmetry, $\sigma_x=0.15$, $\sigma_y=0.85$. This new initial condition leads to another kind of soliton with an elongated shape and two peaks diametrically opposed at the top of the structure. 

\begin{figure}[htbp]
\centering
\includegraphics[width=350pt]{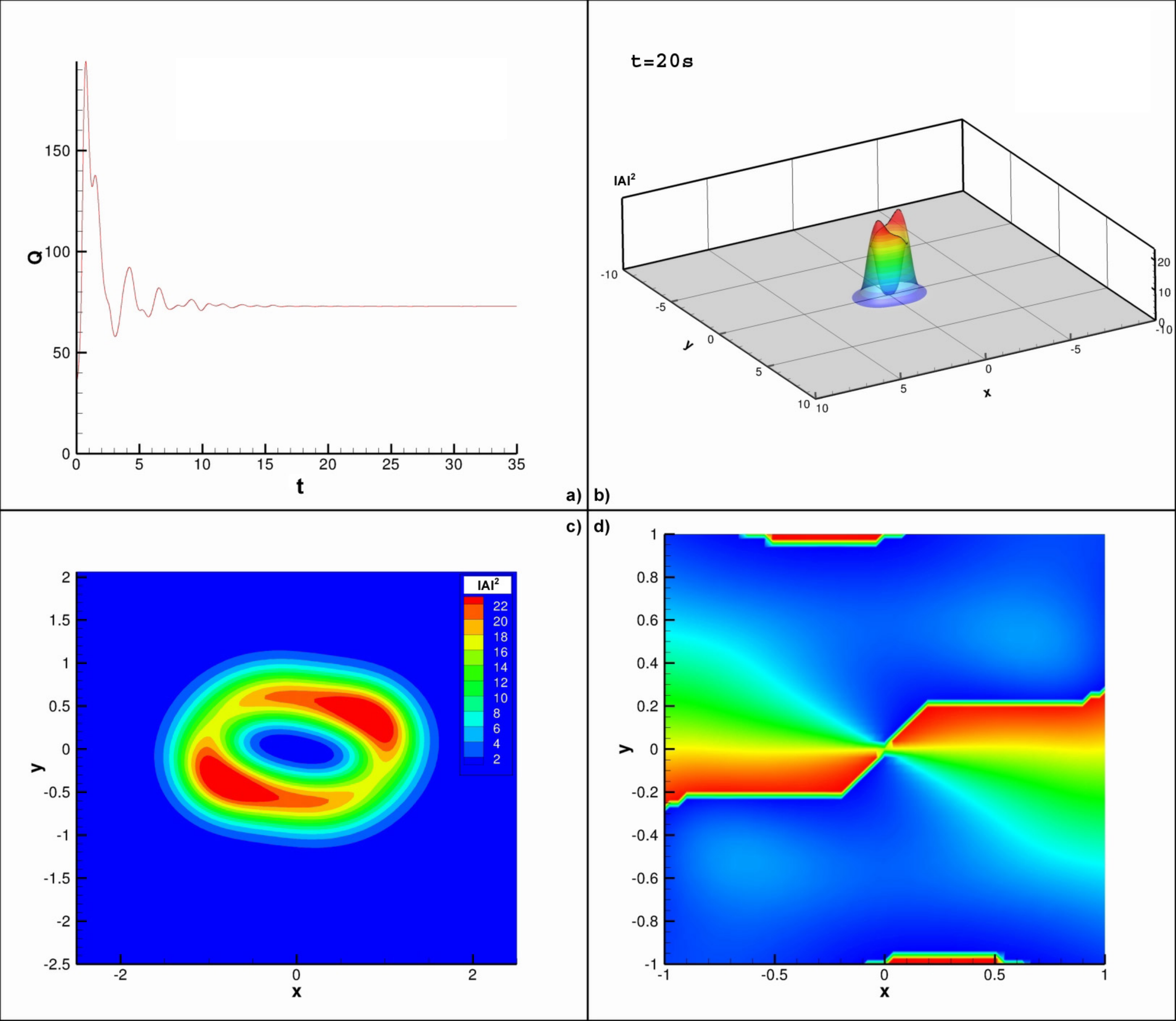}
\caption{Elliptic ring vortex soliton (no radial symmetry). Two peaks appear at the top of the structure, while the phase keeps spinning.  Parameters: line 2  of Table \ref{TabParam}.}
\label{RingEllip}
\end{figure}

The results for both radially symmetric and elliptic solitons are shown in Figs. \ref{Ring} and \ref{RingEllip}.

\begin{figure}[htbp]
\centering
\includegraphics[width=350pt]{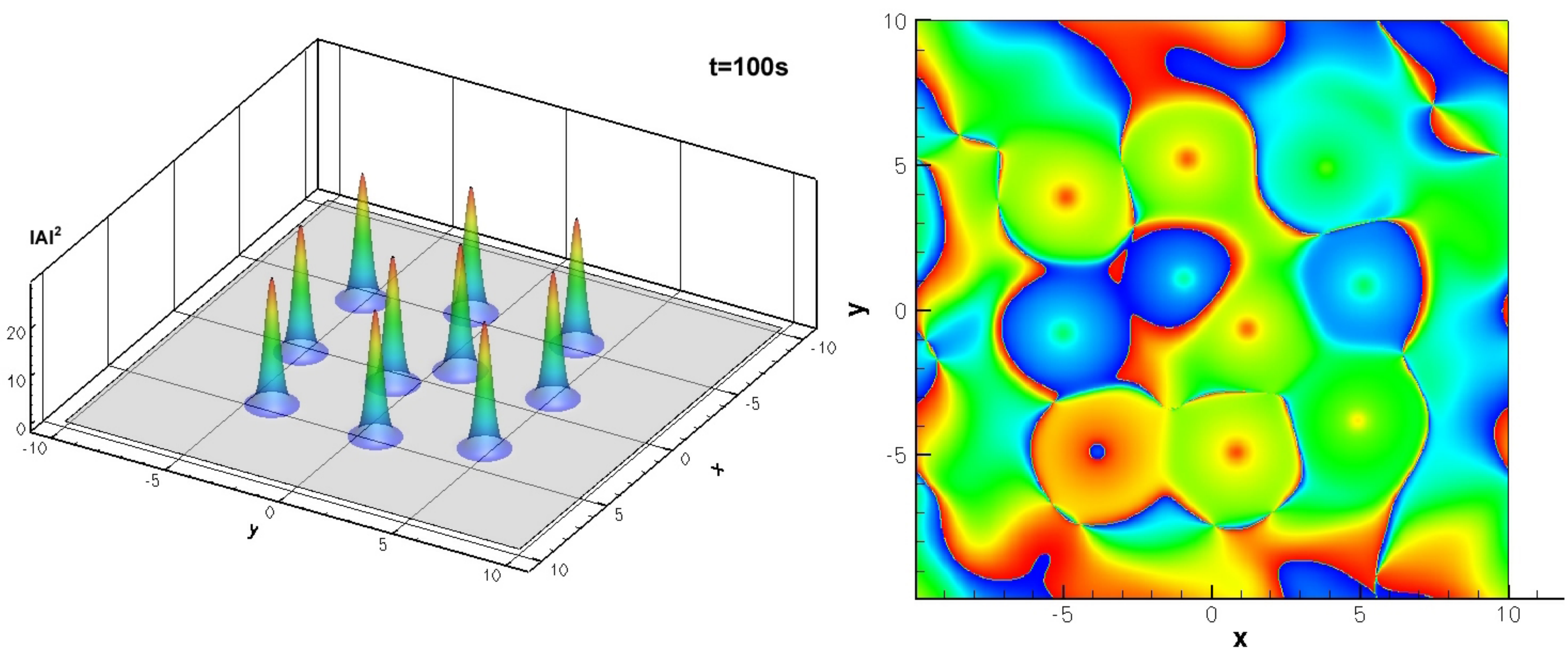}
\caption{Unstable vortex circular ring. Left: 10 bell-shaped solitons appeared in results to the destruction of the initial ring soliton. Right:  the phases of the solitons are not spinning. The parameters of the initial shape are: $A_0=3.0$, $\sigma_x=\sigma_y=0.15$. Parameters: line 2 of Table \ref{TabParam}.}
\label{filament}
\end{figure}

Ordinary bell-shaped solitons with zero-vorticity also exist for the same parameters, but with lower energy. If a ring vortex soliton loses its stability, it will be transformed into several bell-shaped solitons via multiple bifurcations. That is what we have found, by chance, while looking for the pulsating ring vortex soliton, see Fig. \ref{filament}. This class appears to be unstable, may lead to chaos, and needs further investigation.  Stable ring vortices, filamentation and also the unstable ones  using a variational formulation have been found previously by \cite{Skarka:1}. An analysis of 2D doughnut-shaped pulses with phase in the form of a rotating spiral have been found numerically by Crasovan \cite{Crasovan:2}.

\subsection{Pulsating solitons}

The following results contain two interesting features: pulsations and transitions between stable and unstable states. The initial shape is also Gaussian with $A_0=5$, but slightly asymmetric, $\sigma_x=0.8333$ and $\sigma_y=0.9091$. Here we have used the parameters from line 3 of Table \ref{TabParam}. 
\begin{figure}[htbp]
\centering
\includegraphics[width=350pt]{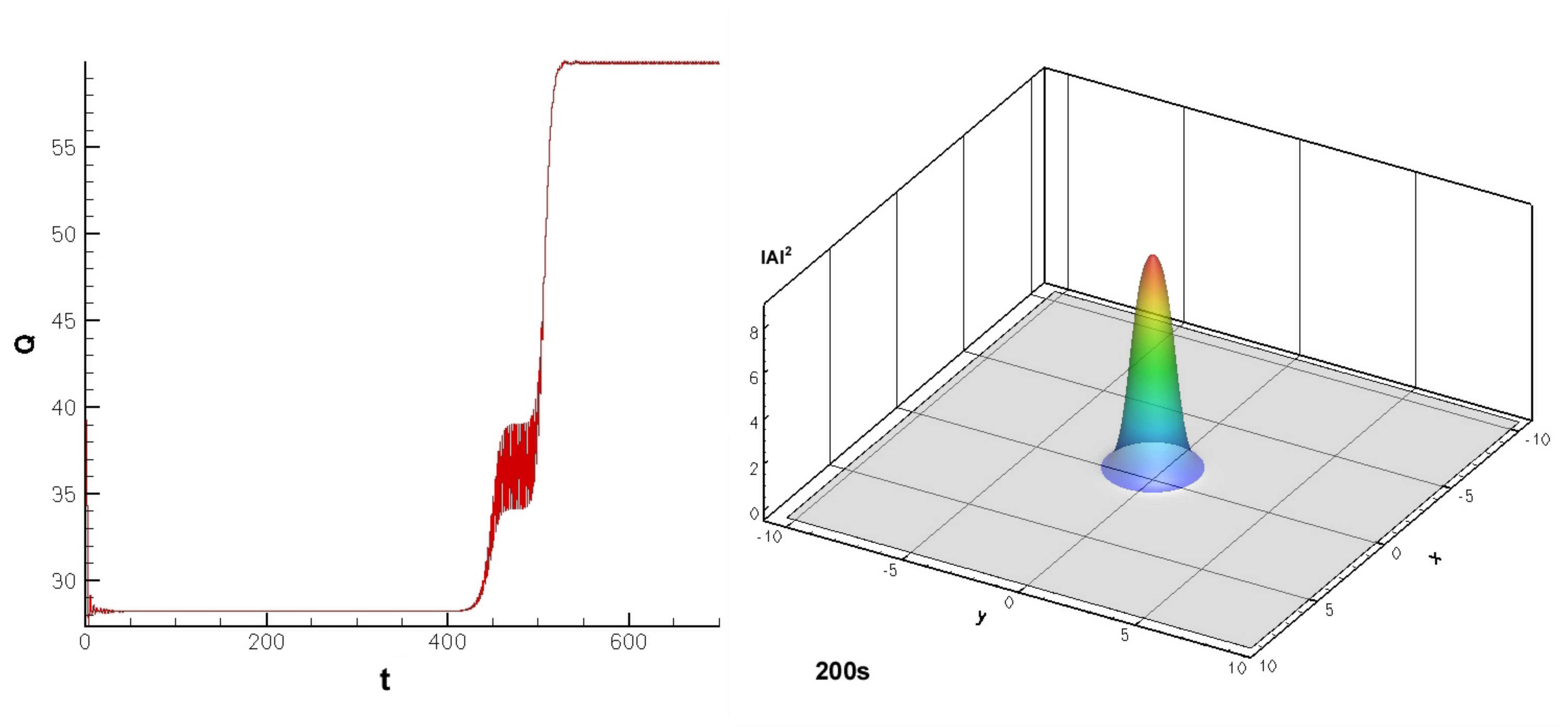}
\caption{Pulsating soliton with changing state. The energy, on the left shows these transitions. Right: bell-shaped soliton $t=200s$.Parameters: line 3 of Table \ref{TabParam}.}
\label{Pulse1}
\end{figure}
\begin{figure}[htbp]
\centering
\includegraphics[width=250pt]{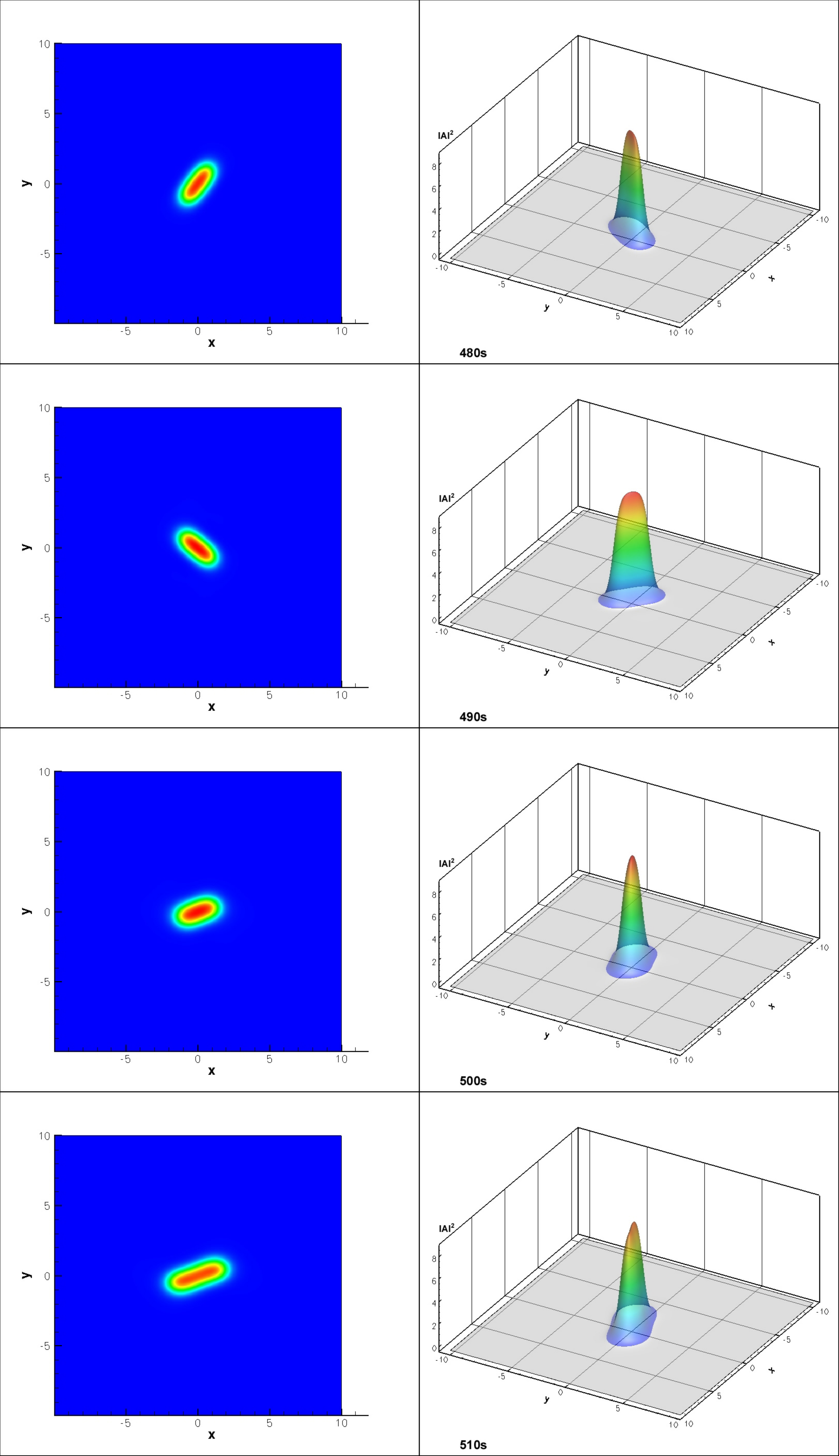}
\caption{During the pulsating state $t \in [480s,520s]$ the soliton elongates alternatively in different directions. Parameters: line 3 of Table \ref{TabParam}.}
\label{Pulse2}
\end{figure}
For the first $200s$, the evolution of the solution looks similar to the stationary soliton, see Fig.\ref{Pulse1}. While the soliton is stabilized at $t=430s$, a perturbation or a bifurcation appears to make the structure evolve toward a new higher state: the energy $Q$ of the soliton is now pulsating. The structure still has radially symmetric profile when $Q$ takes its maximum, but elongates alternatively in different directions when $Q$ takes its minimum value, see Fig.\ref{Pulse2}.

Indeed, this state is not stable for this set of parameters and evolves into another stable soliton which keeps the elongated shape, as if it was a double bell-shaped soliton. The energy still pulsates but with a very little amplitude and the soliton rotates  around its center. Different types of pulsating solitons, with varying energies and shapes, that rotate or not, have been identified, but there is no doubt that there are other types exist too, especially when two or more solitons coexist in the same plane at the same time.

\subsection{Exploding solitons}
\begin{figure}[htbp]
\centering
\includegraphics[width=250pt]{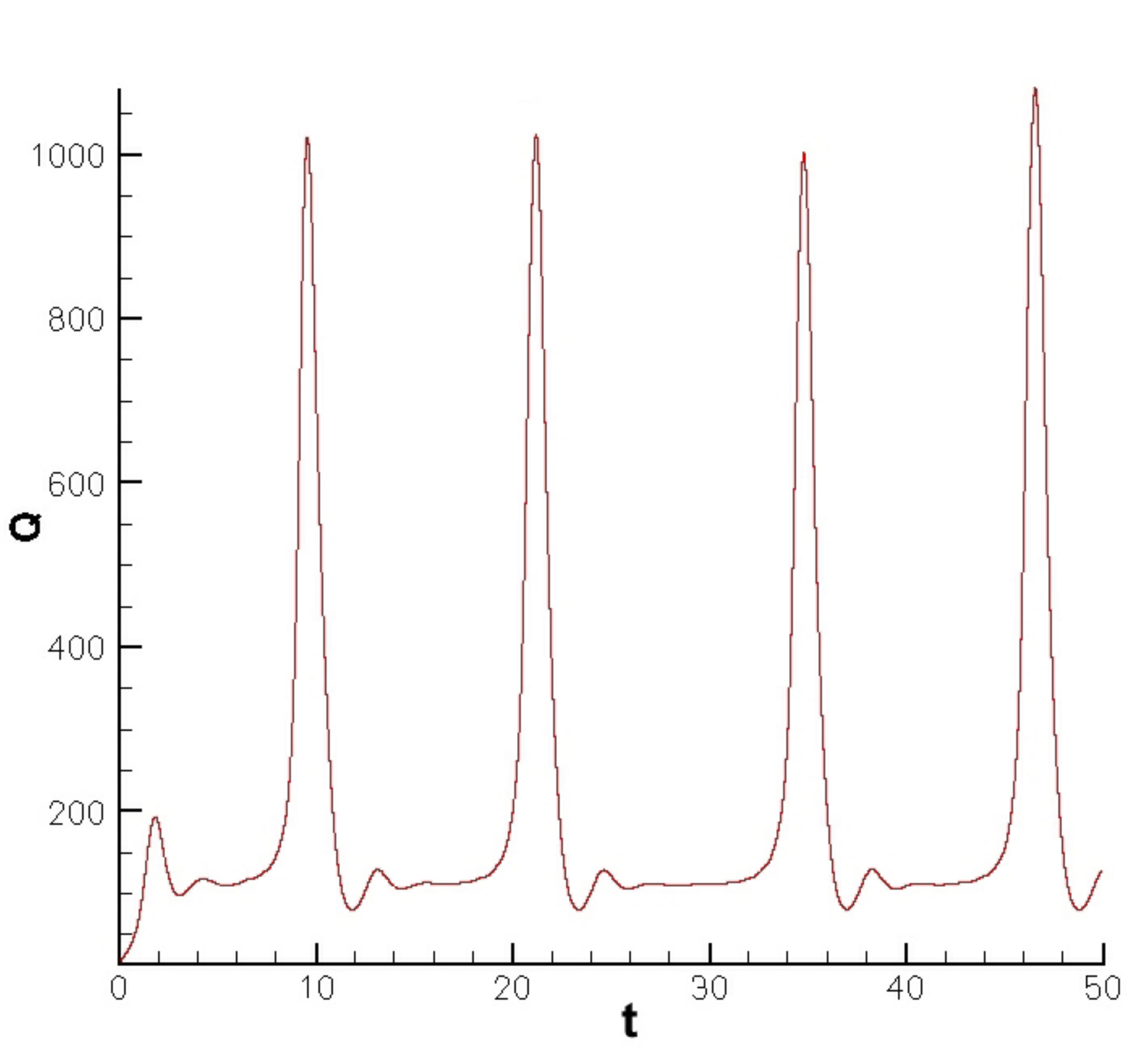}
\caption{Exploding soliton: energy is periodic with high bursts almost every $12s$.  Parameters: line 4 of Table \ref{TabParam}.}
\label{ExplE}
\end{figure}

Another class of 2D solitons is the exploding or erupting solitons. The evolution starts from a localized initial condition of Gaussian shape with parameters $A_0=3.0$, $\sigma_x=\sigma_y=0.3$. After a while its slopes become covered with small scale instabilities which seem to move downwards. The soliton explodes intermittently,
thus resulting in significant bursts of power above the average,
recovering the initial radially cylindrical shape after
each explosion \cite{Ankiewicz:2, Crespo:1}.
Using the same parameters as in the case of the 1D case of papers \cite{Soto:3,Crespo:1}, i.e., $\epsilon=-0.1, b_1=0.125, c_1=0.5, b_3=-1, c_3= 1, b_5=0.1, c_5=-0.6$,  the energy $Q$ for an exploding soliton existing in the 2D case is shown in Fig. \ref{ExplE}. As it was shown for the 1D exploding solitons \cite{Ora:2,Ora:3},
this unusual dynamics appears as a result of an instability.
As we move further from this boundary, the explosions become
more violent, and, as a result more than one beam can
be generated in some cases. This gives rise to a very complicated
dynamics, ending up with the whole numerical grid
filled with the solution. 

\begin{figure}[htbp]
\centering
\includegraphics[width=300pt]{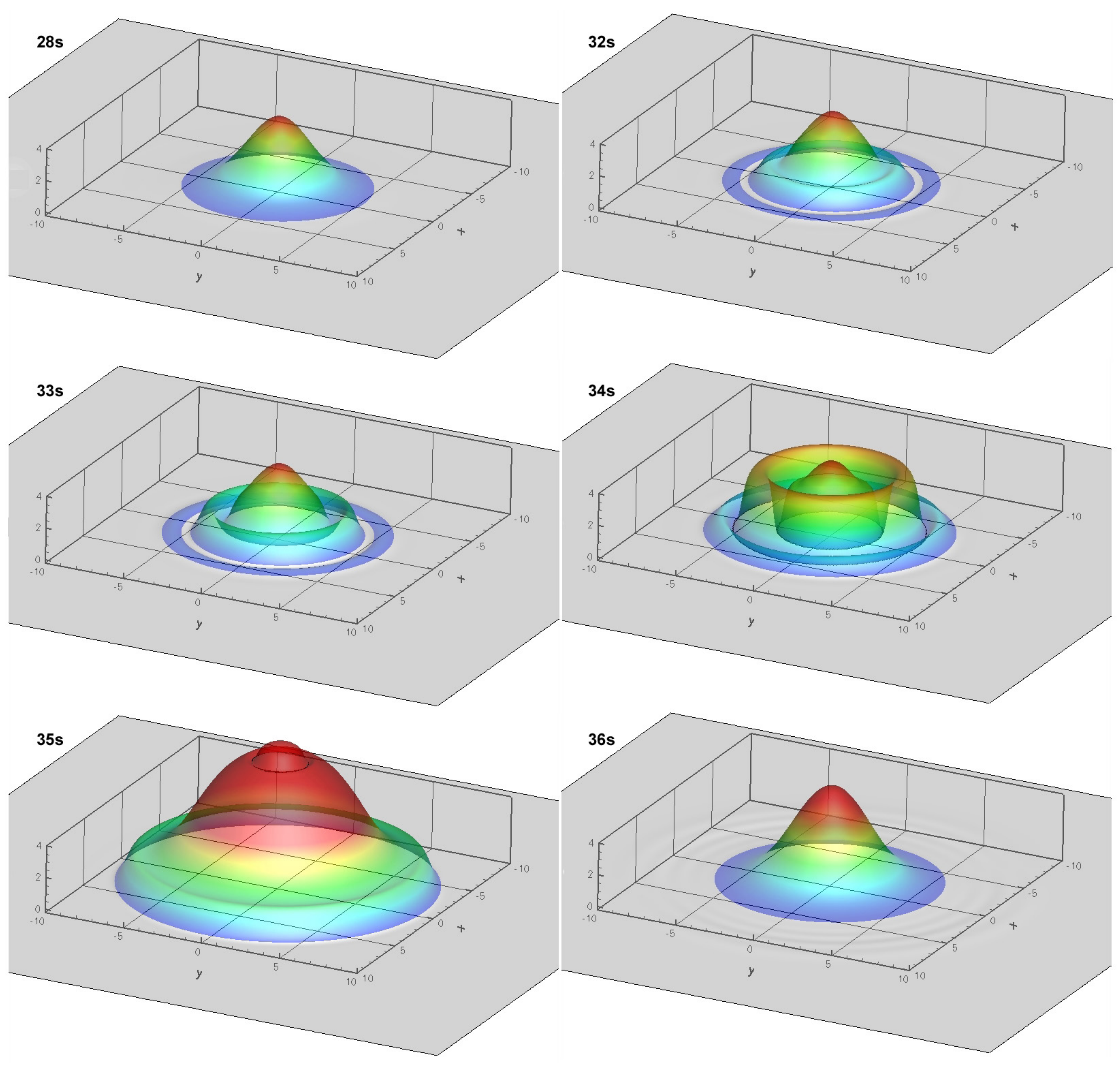}
\caption{Exploding soliton at $t=28s$, the shape is smooth. Then, circular waves appear and grow. Finally, soliton explodes around $t=34s$ and collapses after which it restores back to its initial shape.  Parameters: line 4 of Table \ref{TabParam}.}
\label{ExplA}
\end{figure}

At first, the shape seems to be stable with constant energy and smooth bell-shaped cross section ($t\approx 28s$). Often, circular waves go from the center to the exterior and vanish as if energy were dissipated in the boundaries of the solitons ($t\approx32s$). Suddenly, and periodically, the soliton ``explodes" ($t\approx35s$): its shape grows, so do the energy, until it collapses and goes back to the original shape ($t\approx36s$). This completely chaotic, but well-localized structure restores to a perfect solitons shape after the main burst.  The periodic evolution of the energy, with high bursts of energy can be seen in Fig. \ref{ExplE}. This process never repeats itself in successive periods, however, it always returns to the same shape. Hence, the essential features of the explosions, both observed numerically in 1D  by \cite{Crespo:1,Akhmediev:1} are: explosions occur intermittently, the explosions have similar features but are not identical, and explosions happen spontaneously being triggered by perturbations. Both regular and irregular erupting pulses together with spiraling axisymmetric solitons were also found in \cite{Crasovan:3}. In \cite{Ora:4} authors  have found  the existence of two types of exploding solitons: symmetric and asymmetric. The exploding soliton of Fig. \ref{ExplA}  corresponds to a symmetric transient. This structure will evolve to an asymmetric exploding soliton \cite{Ora:1}.

\section{Conclusion}
In conclusion, based on the numerical simulations, we have presented evidence for the existence of stable stationary 2D solitons in a dissipative medium, described by the 2D CCQGLE, in which asymmetry between space-time variables is included. For the domains of the parameters that we explored, we have also found stable coherent structures in the form of spinning  solitons. These exist for a certain region of parameters, as a result of a competition between focusing nonlinearities and spreading while propagating through medium.  Beyond the regions of stability, the solitons may lose the radial symmetry but remain stable. Varying the parameters, and using Gaussian and ring initial conditions, the simulations revealed more distinct 2D solitons in the form of pulsating, and for the first time two-dimensional exploding solitons with very interesting properties.
\newpage
\bibliographystyle{plain}
\bibliography{bibliography}
\addcontentsline{toc}{section}{References}
\end{document}